\documentclass[twocolumn,floatfix,showpacs,pra,aps]{revtex4}
\usepackage[matrix,frame,arrow]{xypic}
\usepackage{amsfonts,amsmath,amssymb}
\usepackage{graphicx,epsfig} 
 
\begin{document} 
\title{Fault-tolerant quantum computation against biased noise}
\author{Panos Aliferis$\,^1$ and John Preskill$\,^2$}
\affiliation{$^1$IBM T.~J. Watson Research Center, P.~O. Box 218, Yorktown Heights, NY 10598 \\
             $^2$Institute for Quantum Information, California Institute of Technology, Pasadena, CA 91125}
\medskip

\begin{abstract}
We formulate a scheme for fault-tolerant quantum computation that works effectively against highly biased noise, where dephasing is far stronger than all other types of noise. In our scheme, the fundamental operations performed by the quantum computer are single-qubit preparations, single-qubit measurements, and conditional-phase ({\sc cphase}) gates, where the noise in the {\sc cphase} gates is biased. We show that the accuracy threshold for quantum computation can be improved by exploiting this noise asymmetry; {\em e.g.}, if dephasing dominates all other types of noise in the {\sc cphase} gates by four orders of magnitude, we find a rigorous lower bound on the accuracy threshold higher by a factor of five than for the case of unbiased noise.
\end{abstract}

\maketitle


Our confidence that large-scale quantum computers can be realized has been boosted by the theory of fault-tolerant quantum computation \cite{Shor96}, which establishes that noisy quantum computers can operate reliably if the noise is not too strong. In a fault-tolerant simulation of a quantum circuit, logical qubits processed by the computer are protected from damage using a quantum code, and encoded operations are realized by {\em gadgets} constructed from the computer's fundamental operations; aside from performing the desired transformation on the encoded quantum information, gadgets also exploit the redundancy of the code to correct errors caused by the noise.

Most work on fault-tolerant quantum computation has focused on the design of gadgets that work effectively for generic noise without any special structure. But actually, in many physical settings the noise is expected to be highly biased. If the computational basis $\{|0\rangle,|1\rangle\}$ coincides with the energy-eigenstate basis for the unperturbed qubit, then typically dephasing (loss of phase coherence in the computational basis, due to entanglement with the environment) is far stronger than relaxation (bit flips in the computational basis, due to energy exchange with the environment). While dephasing arises from low-frequency noise, relaxation is dominated by noise whose frequency is comparable to the energy splitting. Typically, this higher-frequency noise has a different physical origin than the low-frequency noise responsible for dephasing, and it can be orders of magnitude weaker. In this paper, we analyze fault-tolerant gadgets that are designed to exploit this bias.

The fault-tolerant scheme we propose is founded on the assumption that the quantum-computing hardware can execute a  conditional-phase ({\sc cphase}) gate with highly biased noise, where {\sc cphase} is the diagonal two-qubit gate with eigenvalues $(1,1,1,-1)$ in the computational basis. The complete set of fundamental operations performed by our quantum computer is 
\begin{equation}
\mathcal{G}_{\rm fund}=\{\mathrm{CPHASE}, \mathcal{P}_{|+\rangle}, \mathcal{M}_{\sigma_{\rm x}} \}~ \cup ~\{\mathcal{P}_{|{+}i\rangle}, \mathcal{P}_{|T\rangle}\} \; ,
\end{equation}
\noindent where $\mathcal{M}_{\sigma_{\rm x}}$ denotes the measurement of the Pauli operator $\sigma_{\rm x}$, and $\mathcal{P}_{|\psi\rangle}$ denotes the preparation of a single qubit in the state $|\psi\rangle$. To construct fault-tolerant $\mathcal{G}_{\rm CSS}$ operations (see below), we will need to prepare the state $|+\rangle=\frac{1}{\sqrt{2}}\left(|0\rangle +|1\rangle\right)$, and for  fault-tolerant universal quantum computation, we will also need to prepare the states $|{+}i\rangle = {1\over \sqrt{2}}(|0\rangle + i |1\rangle)$ and $|T\rangle = {1\over \sqrt{2}}(|0\rangle + e^{i\pi/4}|1\rangle)$. We have not listed the identify operation, which is implicitly applied whenever a qubit is idle. 

Our central assumption, that the noise in {\sc cphase} gates is dominated by dephasing, may apply to some proposed gate implementations using semiconductor spins \cite{taylor} and superconducting circuits \cite{koch}; it may also apply to trapped-ion qubits if the {\sc cphase} gates are driven by microwave fields rather than laser pulses \cite{wineland}. 
Furthermore, noise in the preparation of the state $|+\rangle$ is trivially ``biased'' because $|+\rangle$ is an eigenstate of $\sigma_{\rm x}$, and noise in the destructive measurement of $\sigma_{\rm x}$ has no specific structure because the measurement has a classical output. We show that, through appropriate gadget design, this noise bias can be exploited to improve the accuracy threshold for quantum computation; {\em e.g.}, assuming that dephasing dominates all other types of noise by four orders of magnitude, we find that the provable accuracy threshold is higher by more than a factor of five than for the case of unbiased noise.

Other authors \cite{gourlay,stephens,stephens-again} have proposed fault-tolerant gadgets for biased noise, but these previous constructions work only if the noise is dominated by dephasing even for some gates that do not commute with $\sigma_{\rm z}$, such as the controlled-{\sc not} ({\sc cnot}) gate. In our view, this assumption is not physically well motivated. A biased noise model should be applicable if, during the execution of a gate, the perturbation responsible for the noise couples predominantly to the $\sigma_{\rm z}$ components of the qubits. But during the execution of a gate that does not commute with $\sigma_{\rm z}$, the perturbation may not have this property---{\it e.g.}, if the gate is a single-qubit rotation about the $x$ axis, then to take into account a possible over-rotation or under-rotation of the qubit we should include a perturbation proportional to $\sigma_{\rm x}$ rather than $\sigma_{\rm z}$. And even if the perturbation is dominated by a $\sigma_{\rm z}$ term, an insertion of the perturbation {\em during} the execution of a rotation about the $x$ axis will propagate to a linear combination of $\sigma_{\rm z}$ and $\sigma_{\rm y}$, which cannot be described as dephasing noise alone. 
 

Although the biased noise model has a natural formulation in terms of a Hamiltonian that couples the computer to its environment, we will study a {\em stochastic} version of the model. 
A stochastic noise model assigns a probability to each {\em fault path}---{\it i.e.}, to each possible set of faulty fundamental operations in the circuit. We speak of {\em local stochastic noise} with strength $\varepsilon$ if, for any $r$ specified fundamental operations in the circuit, the sum of the probabilities of all fault paths with faults at those $r$ locations is no larger than $\varepsilon^r$ \cite{AGP}. In this model no further restrictions are imposed on the noise and, in particular, the trace-preserving quantum operation applied at the faulty locations of each fault path is arbitrary and can be chosen adversarially. Therefore, although $\varepsilon$ quantifies the strength of the noise, the faults can be correlated both temporally and spatially. It was shown recently in \cite{fibonacci} that an ideal quantum circuit can be simulated accurately and with reasonable overhead provided that $\varepsilon$ is smaller than $\varepsilon_{\rm th}\geq .67\times 10^{-3}$; this rigorous lower bound on the threshold is the best established so far for this noise model. 

The noise model that we will analyze in this paper is a refinement of local stochastic noise that admits two different values of the noise strength: $\varepsilon$, quantifying the rate for faults in preparations and measurements and {\em dephasing} faults in {\sc cphase} gates, and $\varepsilon'\ll \varepsilon$, quantifying the rate for all other types of faults in {\sc cphase} gates. In this model, a fault path indicates not only which locations are faulty, but also, for each {\sc cphase} gate, whether a dephasing fault or some other type of fault has occurred. We speak of {\em local stochastic biased noise} if the sum of the probabilities of all fault paths that are faulty at $r$ specified locations, of which $s$ are non-dephasing faults at {\sc cphase} gates, is no larger than $\varepsilon^{r-s}(\varepsilon')^s$. For dephasing faults, all Kraus operators are assumed to be diagonal in the computational basis, and for all other types of faults, the Kraus operators are arbitrary. We refer to the ratio $\varepsilon/\varepsilon'$ as the noise ``bias.''


Our scheme for fault-tolerant quantum computation will be protected by a code $ \mathcal{C}_1 \triangleright \mathcal{C}_2$ where $\triangleright $ denotes code concatenation. The inner code $\mathcal{C}_1$ protects against dephasing, mapping highly biased noise to a more balanced {\em effective} noise model with reduced noise strength. The code $\mathcal{C}_2$ maps unstructured noise with strength below $\varepsilon_{\rm th}$ to noise with negligible strength. We take $\mathcal{C}_1$ to be a length-$n$ repetition code in the dual basis, where $n$ is odd; the $n{-}1$ check operators are $I^{\otimes j} \otimes \sigma_{\rm x} \otimes \sigma_{\rm x} \otimes I^{\otimes n-j-2}$ ($j = 0,1,\dots, n-2$), and the logical Pauli operators acting on the one encoded qubit are $\sigma_{\rm x}^L=\sigma_{\rm x}\otimes I\otimes I\otimes \dots \otimes I$, and $\sigma_{\rm z}^L=\sigma_{\rm z} \otimes \sigma_{\rm z} \otimes \dots  \otimes \sigma_{\rm z}$. The code $\mathcal{C}_1$ can correct  $(n-1) / 2 $  $\sigma_{\rm z}$ errors but provides no protection against $\sigma_{\rm x}$ errors. We take $\mathcal{C}_2$ to be a concatenated CSS code. For a CSS code \cite{AGP}, the fault-tolerant encoded versions of operations in the set
\begin{equation}
\mathcal{G}_{\rm CSS}=\{\mathrm{CNOT}, \mathcal{P}_{|0\rangle}, \mathcal{P}_{|+\rangle}, \mathcal{M}_{\sigma_{\rm z}}, \mathcal{M}_{\sigma_{\rm x}} \} \; 
\end{equation}
can be built out of operations that are contained in this set; furthermore, $\mathcal{G}_{\rm CSS}$ operations suffice for measuring the error syndrome. 

\begin{figure}[tb]
\begin{center}
\begin{tabular}{c}
\put(-5.1,0){\includegraphics[width=16cm,keepaspectratio]{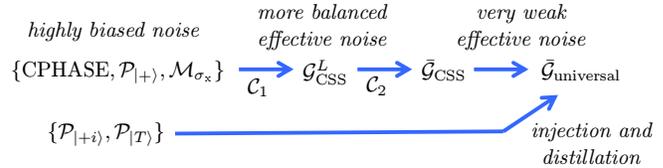}}
\vspace{-10.1cm}
\end{tabular}
\end{center}\caption{Scheme for achieving fault-tolerant universal quantum computation against highly biased noise.}
\label{fig:scheme}
\end{figure}

We will use the fundamental operations in $\mathcal{G}_{\rm fund}$ (where the only necessary state preparation is $\mathcal{P}_{|+\rangle}$) to construct $\mathcal{G}_{\rm CSS}$ gadgets protected by $\mathcal{C}_1$. Combining with known constructions for CSS codes \cite{AGP2}, we obtain $\mathcal{G}_{\rm CSS}$ gadgets protected by $\mathcal{C}_1 \triangleright  \mathcal{C}_2$. Finally, CSS operations will be extended to a universal set by appending preparations of the states $|{+}i\rangle$ and $|T\rangle$; high fidelity encoded copies of these states can be prepared by teleporting (``injecting'') into the $\mathcal{C}_1 \triangleright  \mathcal{C}_2$ block and then performing state distillation \cite{bravyi}. Our scheme for achieving fault-tolerant universal quantum computation is illustrated in Fig.~\ref{fig:scheme} (we denote an operation $\mathcal{O}$ encoded in $\mathcal{C}_1$ or $\mathcal{C}_1 \triangleright \mathcal{C}_2$ as $\mathcal{O}^L$ or $\mathcal{\bar O}$ respectively).

The crux of our gadget constructions, and the basis for our threshold analysis, is the design of the $\mathcal{C}_1$-protected {\sc cnot} gadget using the operations in $\mathcal{G}_{\rm fund}$. The key idea is to use a variant of state teleportation that simultaneously executes the encoded gate and extracts the error syndrome. But first, let us discuss how to construct $\mathcal{C}_1$-protected gadgets for the other operations in $\mathcal{G}_{\rm CSS}$.

The destructive measurement of $\sigma_{\rm x}^L$ is performed by measuring $\sigma_{\rm x}$ for all qubits in the code block, and then computing the majority of the measurement outcomes. This measurement is fault tolerant in the following sense: if $m$ of the qubits in the input code block have errors, and $s$ of the single-qubit measurements are faulty, then the outcome of the noisy encoded measurement agrees with the outcome of an ideal encoded measurement provided that $m{+}s\le (n{-}1)/2$. The preparation of  $|+\rangle_L$ is executed transversally: $\mathcal{P}_{|+\rangle_L}=\left(\mathcal{P}_{|+\rangle}\right)^{\otimes n}$. This operation is fault tolerant because at least $(n{+}1)/2$ of the preparations of $|+\rangle$ must be faulty to cause an encoded error.

\begin{figure}[tb]
\begin{center}
\begin{tabular}{c}
\put(-4,0){\includegraphics[width=16cm,keepaspectratio]{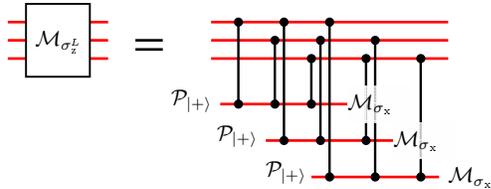}}
\vspace{-9.8cm}
\end{tabular}
\end{center}\caption{Gadget that measures $\sigma_{\rm z}^L$ (here, $n=3$). The measurement is repeated $r$ times to ensure fault tolerance (here, $r=3$), with the repetitions staggered as shown so that the data qubits are never idle in between consecutive gates.}
\label{fig:z-measurement}
\end{figure}

A nondestructive measurement of $\sigma_{\rm z}^L$ is performed with the circuit depicted in Fig.~\ref{fig:z-measurement}: an ancilla qubit is prepared in the state $|+\rangle$, $n$ consecutive {\sc cphase} gates are executed, and then $\sigma_{\rm x}$ is measured on the ancilla qubit. If performed only once, this measurement is not fault tolerant, because a single $\sigma_{\rm z}$ error acting on the ancilla can flip the outcome; however, fault tolerance can be ensured by repeating the measurement $r$ times, where $r$ is odd, and computing the majority of the outcomes. Although $\sigma_{\rm z}$ errors acting on the data qubits do not affect the measurement outcome, they might contribute to a logical $\sigma_{\rm z}^L$ error that could affect subsequent operations. It is therefore noteworthy that if the input block has $m$ $\sigma_{\rm z}$ errors and the measurement gadget has $s$ dephasing faults, then there will be no more than $m{+}s$ $\sigma_{\rm z}$ errors in the output block. 

The preparation of $|0\rangle_L$ is executed by first preparing $|+\rangle_L$, and then performing the nondestructive measurement of $\sigma_{\rm z}^L$. Again, $\sigma_{\rm z}$ errors acting on the data qubits will not disturb the eigenvalue of $\sigma_{\rm z}^L$; only the $\sigma_{\rm z}$ errors acting on the ancilla are problematic. Therefore, the fault tolerance of $\mathcal{M}_{\sigma_{\rm z}^L}$ ensures the fault tolerance of $\mathcal{P}_{|0\rangle_L}$. If the measurement result is $\sigma_{\rm z}^L=-1$, then the prepared state differs from $|0\rangle_L$ by a known logical $\sigma_{\rm x}^L$ error. This error need not be corrected; rather it is used to update the ``Pauli frame'' of the encoded block \cite{knill}. 

\begin{figure}[tb]
\begin{center}
\begin{tabular}{c}
\put(-4.3,0){\includegraphics[width=16cm,keepaspectratio]{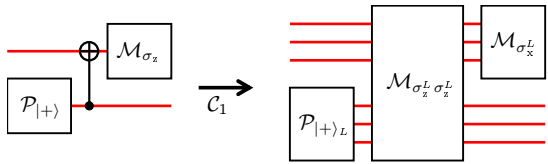}}
\vspace{-10.1cm}
\end{tabular}
\end{center}\caption{On the left, a ``one-bit teleportation'' circuit. On the right, the error correction gadget. 
                    }
\label{fig:one-bit}
\end{figure}

Error correction can be performed by teleporting an encoded block \cite{knill}. Because we only need to correct $\sigma_{\rm z}$ errors, the error correction gadget can be simplified to an encoded version of the ``one-bit teleportation'' circuit \cite{one-bit} depicted in Fig.~\ref{fig:one-bit}. Ideally, the output encoded qubit has the same state as the input encoded qubit, apart from a possible $\sigma_{\rm z}^L$ (if the outcome of $\mathcal{M}_{\sigma_{\rm x}^L}$ is ${-}1$) and a possible $\sigma_{\rm x}^L$ (if the outcome of $\mathcal{M}_{\sigma_{\rm z}^L\sigma_{\rm z}^L}$ is ${-}1$); thus the Pauli frame is updated based on the measurement outcomes. The nondestructive measurement of $\sigma_{\rm z}^L\sigma_{\rm z}^L$ is performed using one ancilla qubit and $2n$ {\sc cphase} gates, and is repeated $r$ times, much as for the measurement of $\sigma_{\rm z}^L$ described above. If the input block has $m$ $\sigma_{\rm z}$ errors and the error-correction gadget has $s$ dephasing faults, then the outcome of $\mathcal{M}_{\sigma_{\rm x}^L}$ agrees with the ideal case, provided $m{+}s\le (n{-}1)/2$. Furthermore the outcome of $\mathcal{M}_{\sigma_{\rm z}^L\sigma_{\rm z}^L}$ agrees with an ideal measurement for $s \le (r{-}1)/2$, and the number of $\sigma_{\rm z}$ errors in the output block is no more than $s$.

\begin{figure}[tb]
\begin{center}
\begin{tabular}{c}
\leavevmode
\epsfxsize=6in
\put(-4.5,0){\includegraphics[width=17.5cm,keepaspectratio]{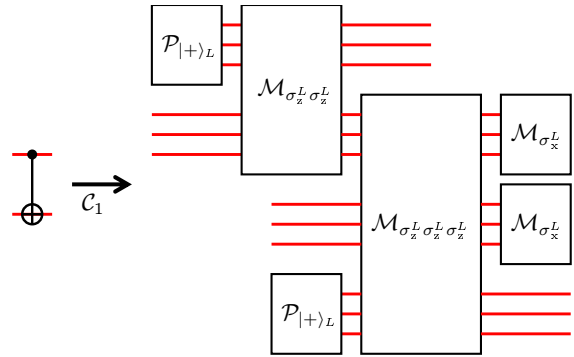}}
\vspace{-8.6cm}
\end{tabular}
\end{center}\caption{Fault-tolerant {\sc cnot} gadget. Pauli operators that update the Pauli frame (not shown) are determined by the measurement outcomes. 
           }
\label{fig:cnot}
\end{figure}

By combining one-bit teleportation gadgets acting on both output blocks with a logical {\sc cnot} gate, we obtain the {\sc cnot} gadget depicted in Fig.~\ref{fig:cnot}, where the upper block is the control block and the lower block is the target block. The measurements of $\sigma_{\rm z}^L\sigma_{\rm z}^L$ and $\sigma_{\rm z}^L\sigma_{\rm z}^L\sigma_{\rm z}^L$ are repeated $r$ times using ancilla qubits and {\sc cphase} gates, as described previously. If the input target block has $m_1$ $\sigma_{\rm z}$ errors, the input control block has $m_2$ $\sigma_{\rm z}$ errors, and the {\sc cnot} gadget contains $s$ dephasing faults, then both $\mathcal{M}_{\sigma_{\rm x}^L}$'s agree with the ideal case provided that $m_1{+}s\le (n{-}1)/2$ and $m_2{+}s\le (n{-}1)/2$; furthermore, each output block contains no more than $s$ $\sigma_{\rm z}$ errors, and $\mathcal{M}_{\sigma_{\rm z}^L\sigma_{\rm z}^L}$ and $\mathcal{M}_{\sigma_{\rm z}^L\sigma_{\rm z}^L\sigma_{\rm z}^L}$ agree with the ideal measurements for $s\le (r{-}1)/2$. Further properties of the {\sc cnot} gadget are discussed in Appendix A.

Now consider a circuit constructed from these $\mathcal{C}_1$-protected $\mathcal{G}_{\rm CSS}$ gadgets---how accurately does it simulate an ideal circuit? A gadget operates correctly if all of the encoded measurements it contains agree with the ideal case (the case in which the input blocks have no errors and the gadget contains no faults); otherwise the gadget fails. For each $\mathcal{G}_{\rm CSS}$ gadget, we have derived an upper bound on its probability of failure in terms of the noise strength $\varepsilon$ and the bias factor $\varepsilon/\varepsilon'$ of the local stochastic biased noise model. See Appendix B for details of this combinatorial analysis. 

\begin{figure}[tb]
\begin{center}
\begin{tabular}{c}
\put(-4.5,0){\includegraphics[width=9.3cm,keepaspectratio]{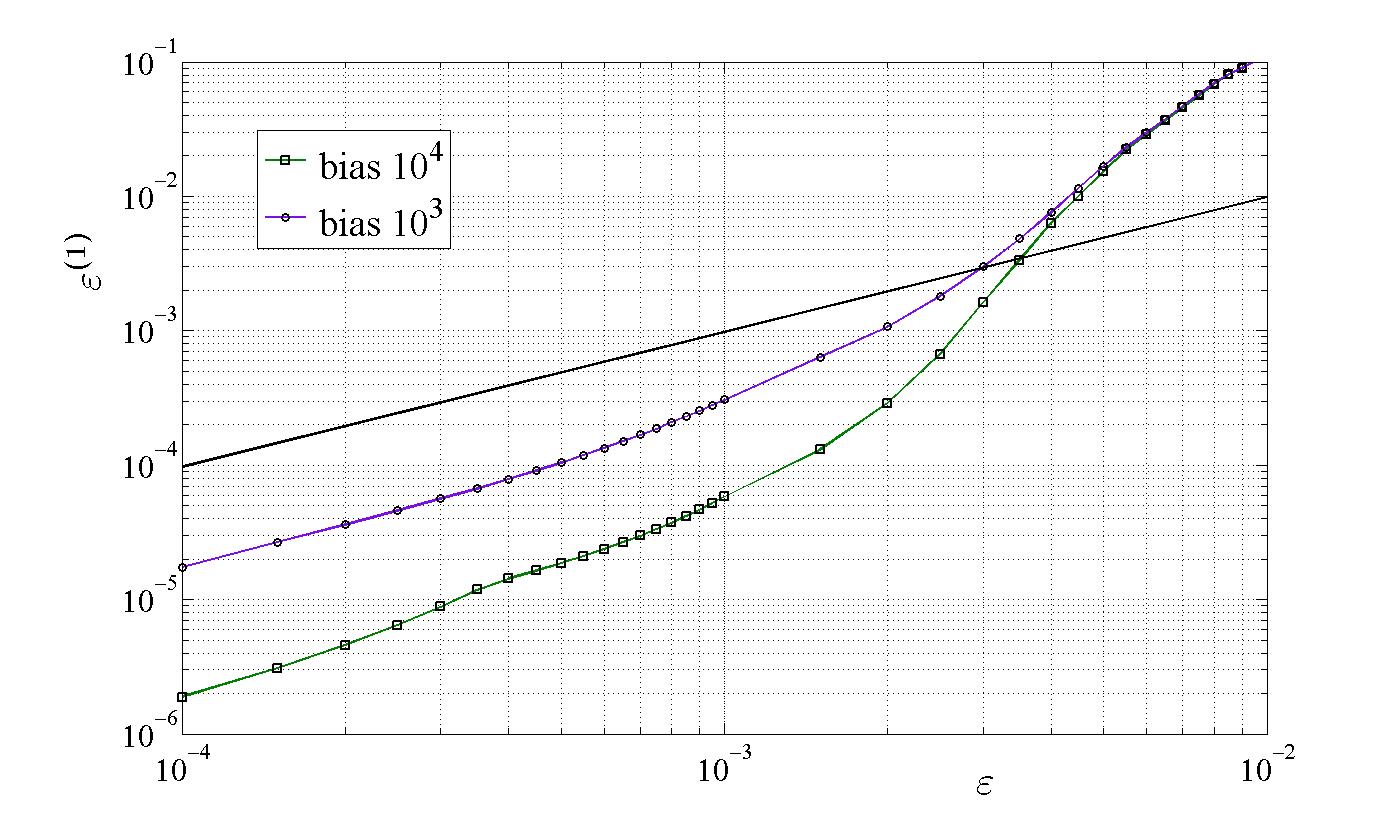}}
\vspace{-0.9cm}
\end{tabular}
\end{center}\caption{Upper bounds on the effective noise strength $\varepsilon^{(1)}$ for $\mathcal{C}_1$-protected $\mathcal{G}_{\rm CSS}$ operations as a function of $\varepsilon$ and the bias $\varepsilon/\varepsilon'$ (where for each value of $\varepsilon$, we optimize over $n$ and $r$). The straight line with slope unity serves as a guide to the eye.}
\label{fig:plot}
\end{figure}

The largest of these upper bounds (found for the {\sc cnot} gadget) is denoted $\varepsilon^{(1)}$; it can be regarded as the effective noise strength of a local stochastic noise model that characterizes the noise in $\mathcal{C}_1$-protected $\mathcal{G}_{\rm CSS}$ circuits. In Fig.~\ref{fig:plot}, we have plotted $\varepsilon^{(1)}$ as a function of $\varepsilon$ for two different values of the bias.  If $\varepsilon^{(1)}$ is below the previously established lower bound on the accuracy threshold for CSS operations ($\varepsilon_{\rm th}^{\rm CSS}\ge .67\times 10^{-3}$ \cite{fibonacci}), then we can choose the code $\mathcal{C}_2$ so that the $\mathcal{G}_{\rm CSS}$ gadgets protected by $\mathcal{C}_1 \triangleright  \mathcal{C}_2$ are arbitrarily accurate. Thus we set $\varepsilon^{(1)}=.67\times 10^{-3}$, and choose $n$ so that $\varepsilon$ is as large as possible.  
If the bias is $10^{4}$, then the maximum value is $\varepsilon=2.50\times 10^{-3}$, occurring at $n=r=11$. 

Furthermore, as we show in Appendix C, the injection and distillation of the $|{+}i\rangle$ and $|T\rangle$ states can be performed effectively for $\varepsilon \leq 2.50\times 10^{-3}$. We conclude that $2.50\times 10^{-3}$ is a lower bound on the accuracy threshold for universal quantum computation under local stochastic biased noise with bias $10^{4}$. This is an improvement by about a factor of four compared to the case of unstructured noise. On the other hand, the improvement is more modest for smaller values of bias; {\em e.g.}, for a bias of $10^3$, our lower bound becomes $1.54\times 10^{-3}$ (where the maximum value $\varepsilon=1.54\times 10^{-3}$ occurs at $n=r=7$), an improvement by about a factor of two.

These results can be further improved by modifying the decoding procedure for $\mathcal{C}_1 \triangleright \mathcal{C}_2$. Recall that the outcome of each $\mathcal{C}_1$-protected measurement is determined by a majority vote, and note that this outcome is more likely to be wrong if the vote is ``close''---{\em i.e.}, if the majority has just one more vote than the minority. For example, if $(n{-1})/2$ of the ${\sigma_x}$ measurements inside $\mathcal{M}_{\sigma_{x}^L}$ disagree with the choice of the majority, then there might be $(n{+1})/2$ errors in the block, resulting in an encoded error. But if only $(n{-}3)/2$ qubits disagree with the majority, then there must be at least $(n{+}3)/2$ errors to cause an encoded error. This observation also applies to the majority voting in $\mathcal{M}_{\sigma_{\rm z}^L\sigma_{\rm z}^L}$ and $\mathcal{M}_{\sigma_{\rm z}^L\sigma_{\rm z}^L\sigma_{\rm z}^L}$. 

The code $\mathcal{C}_2$ can be decoded more reliably by exploiting information concerning which $\mathcal{C}_1$-protected measurements have close votes; see Appendix D for details. Using this more sophisticated decoding method, we find that the accuracy threshold improves to $2.09\times 10^{-3}$ for bias $10^3$, and improves to $3.51\times 10^{-3}$ for bias $10^{4}$.


Fault-tolerant methods will be essential for achieving large-scale quantum computation. These methods can be more effective when customized for the particular properties of the noise in the computing hardware. In this paper, we have explained how to exploit noise asymmetry in diagonal gates to make fault-tolerant quantum computing work better. We have analyzed the performance of our scheme for local stochastic biased noise; using techniques described in \cite{terhal,AGP,AKP}, a more realistic Hamiltonian biased noise model could also be analyzed. 

A notable property of our constructions is that the only fundamental operation used by the quantum computer, other than single-qubit preparations and measurements, is the two-qubit {\sc cphase} gate. This feature is attractive, because in some physical settings {\sc cphase} gates are particularly easy to execute with highly biased noise and reasonable fidelity; {\em e.g.}, our companion paper \cite{koch2} discusses how the encoding scheme we have formulated here applies to superconducting flux qubits.  

 
\medskip
We thank David DiVincenzo, Daniel Gottesman, and Gabriel Mendoza for useful discussions. This research is supported in part by DoE under Grant No. DE-FG03-92-ER40701, NSF under Grant No. PHY-0456720, and NSA under ARO Contract No. W911NF-05-1-0294.

\bibliographystyle{unsrt}


\appendix

\cleardoublepage



\section{Comments on the {\sc cnot} gadget}

In our discussion in the main text, the measurement of $\sigma_{\rm z}^L\sigma_{\rm z}^L$ is repeated $r$ times inside $\mathcal{M}_{\sigma_{\rm z}^L\sigma_{\rm z}^L}$ and the measurement of $\sigma_{\rm z}^L\sigma_{\rm z}^L\sigma_{\rm z}^L$ is also repeated $r$ times inside $\mathcal{M}_{\sigma_{\rm z}^L\sigma_{\rm z}^L\sigma_{\rm z}^L}$. In general, the number of repetitions could be different in the two cases, $r_1$ and $r_2$ for the measurements of $\sigma_{\rm z}^L\sigma_{\rm z}^L$ and $\sigma_{\rm z}^L\sigma_{\rm z}^L\sigma_{\rm z}^L$ respectively, and for now we will distinguish $r_1$ and $r_2$ from $n$ so that the counting we describe below will be more transparent. In fact, later on we will set $r_1=r_2=n$, which turns out to be optimal or nearly optimal in the cases we have studied. A further advantage of the choice $r_1=r_2=n$ is that we can eliminate storage locations (where qubits are idle) in the {\sc cnot} gadget by staggering the measurements as in Fig.~\ref{fig:z-measurement}; for this reason we will not include any faults at storage locations in our estimate of the failure probability. 

\begin{figure}[tb]
\begin{center}
\begin{tabular}{c}
\put(-3.5,0){\includegraphics[width=15cm,keepaspectratio]{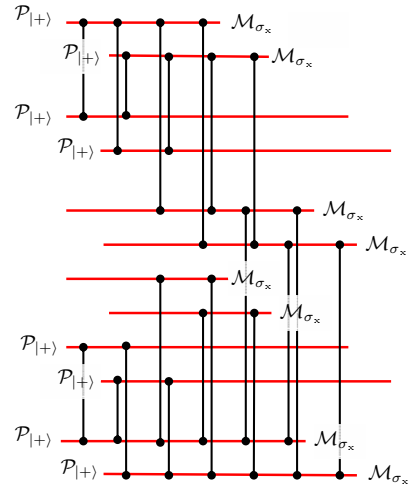}}
\vspace{-4.8cm}
\end{tabular}
\end{center}\caption{The full circuit of the {\sc cnot} gadget (here, for compactness we have chosen $r_1=r_2=n=2$, but the structure of the circuit is similar for odd values).
                    }
\label{fig:explicit-cnot}
\end{figure}

A noteworthy property of the {\sc cnot} gadget is that, if $r_1=r_2=n$, and if the measurements are staggered as in Fig.~\ref{fig:z-measurement}, then the latest operations on the output blocks act one time step {\em before} the earliest operations on the input blocks. This property is obscured by the diagrammatic notation in Fig.~\ref{fig:cnot}, but it is evident once we consider the full circuit as in Fig.~\ref{fig:explicit-cnot}. Let us say that a data qubit ``interacts'' in a time step in which it is coupled to an ancilla qubit by a {\sc cphase} gate. We choose a standard ordering for the $n$ qubits in each block, such that the interactions of qubit $j$ lag $j{-}1$ time steps behind the interactions of qubit 1. Then, in the {\sc cnot} gadget, qubit 1 in the output control block interacts during time steps 1 through $n$, and qubit 1 of the output target block interacts during time steps 1 through $n$. Meanwhile, qubit 1 in the input control block interacts during time steps $n{+}1$ through $3n$, and qubit 1 in the input target block interacts during time steps $n{+}1$ through $2n$. Therefore, in time step $n{+}1$, as qubit 1 in each input block begins to interact, qubit 1 in each output block is ready for execution of the next gadget. This is a characteristic feature of gate teleportation; it implies that a circuit of $\mathcal{G}_{\rm CSS}$ operations can be simulated in constant depth, independent of the size and depth of the simulated circuit.

\section{The threshold for $\mathcal{G}_{\rm CSS}$ operations}

Among all $\mathcal{G}_{\rm CSS}$ gadgets, the {\sc cnot} gadget contains the largest number of fundamental operations. Therefore, if we derive an upper bound on its probability of failure, this bound will also apply to all the other $\mathcal{G}_{\rm CSS}$ gadgets.

To estimate the probability of failure for a {\sc cnot} gadget, we first observe that a $\sigma_{\rm x}$ error on a data qubit or $\sigma_{\rm z}$ errors acting on many ancilla qubits can cause an incorrect outcome of $\mathcal{M}_{\sigma_{\rm z}^L\sigma_{\rm z}^L}$ or $\mathcal{M}_{\sigma_{\rm z}^L\sigma_{\rm z}^L\sigma_{\rm z}^L}$.  The outcome of $\mathcal{M}_{\sigma_{\rm x}^L}$ may be incorrect due to $\sigma_{\rm z}$ errors acting on multiple qubits in a single block, or due to a $\sigma_{\rm x}$ error acting on an ancilla qubit that propagates repeatedly to generate many $\sigma_{\rm z}$ errors in the code block.

\begin{figure*}[tb]
\begin{center}
\begin{tabular}{c}
\put(-4,-2.5){\includegraphics[width=13cm,keepaspectratio]{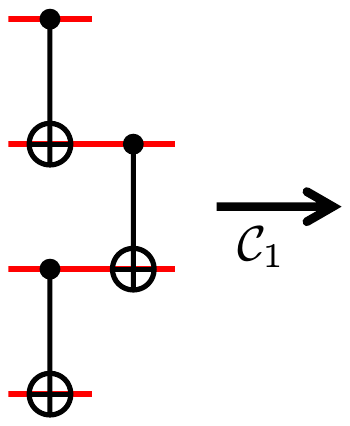}}
\put(-1.8,0){\includegraphics[width=13cm,keepaspectratio]{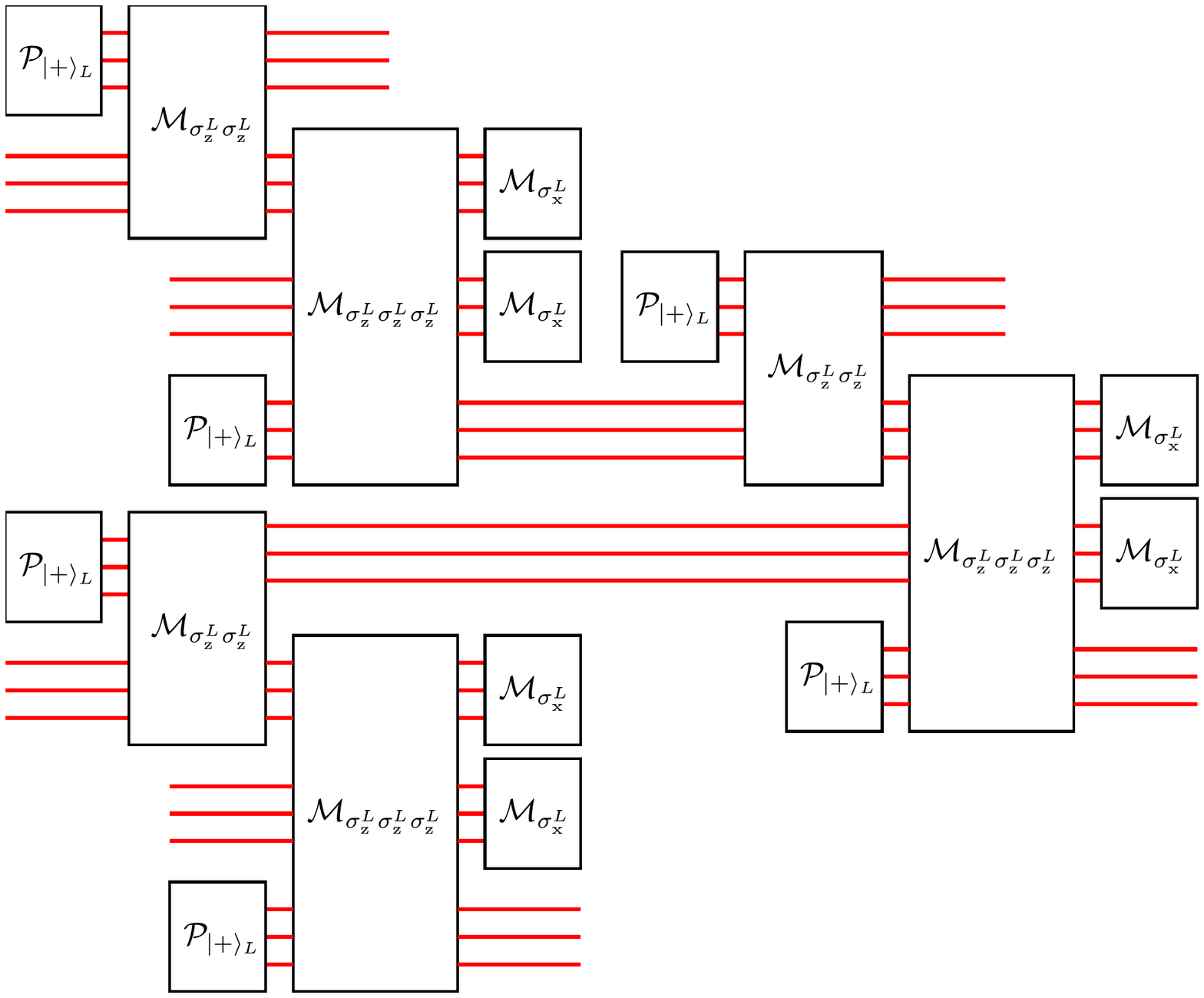}}
\vspace{-5.2cm}
\end{tabular}
\end{center}\caption{A {\sc cnot} gadget preceded by {\sc cnot} gadgets acting on each of its input blocks.}
\label{fig:three-cnots}
\end{figure*}

In addition, we must take into account errors in preceding gadgets that could propagate into the {\sc cnot} gadget we are considering; {\em e.g.}, Fig.~\ref{fig:three-cnots} depicts a {\sc cnot} gadget preceded by {\sc cnot} gadgets acting on each of its input blocks. A $\sigma_{\rm z}$ error in one of the preceding gadgets might affect the outcome of a $\mathcal{M}_{\sigma_{\rm x}}^L$ in the current gadget, and a $\sigma_{\rm x}$ error in one of the preceding gadgets can alter the outcome of $\mathcal{M}_{\sigma_{\rm z}^L\sigma_{\rm z}^L}$ or $\mathcal{M}_{\sigma_{\rm z}^L\sigma_{\rm z}^L\sigma_{\rm z}^L}$ in the current gadget.

To understand the effect of $\sigma_{\rm x}$ errors in the {\sc cnot} gadgets, note that an incorrect outcome of $\mathcal{M}_{\sigma_{\rm z}^L\sigma_{\rm z}^L}$ produces a $\sigma_{\rm x}^L$ error acting on the {\sc cnot} gadget's output control block, and an incorrect outcome of $\mathcal{M}_{\sigma_{\rm z}^L\sigma_{\rm z}^L\sigma_{\rm z}^L}$ produces a  $\sigma_{\rm x}^L$ error acting on the {\sc cnot} gadget's output target block. Consider, {\em e.g.}, the control block of the final {\sc cnot} gadget for the case depicted in Fig.~\ref{fig:three-cnots}, and suppose that a single non-dephasing fault in a {\sc cphase} gate contained in the $\mathcal{M}_{\sigma_{\rm z}^L\sigma_{\rm z}^L\sigma_{\rm z}^L}$ of the immediately preceding gadget alters the outcome of that measurement and also of the $\mathcal{M}_{\sigma_{\rm z}^L\sigma_{\rm z}^L}$ in the later gadget. Then this one fault causes logical errors in each of two consecutive gadgets. However, the logical error in the earlier {\sc cnot} gadget is a $\sigma_{\rm x}^L$ acting on its output target block, and has no effect on its output control block. Therefore, we can propagate the logical error forward from the earlier gadget to the later gadget; {\em i.e.}, we may just as well say that the earlier gadget is executed properly, and the logical error occurs only in the later gadget.

More generally, whenever a single $\sigma_{\rm x}$ error causes logical errors in two consecutive gadgets, the logical error in the earlier gadget can be propagated forward into the later gadget in this way. Thus, we may hold the  $\sigma_{\rm x}$ error responsible for only the failure of the later gadget, and we may disregard the damage it inflicts on the earlier gadget. In effect, then, a single non-dephasing fault occurring with probability $\varepsilon'$ can cause the failure of only one of our $\mathcal{C}_1$-protected gadgets.

A measurement of $\sigma_{\rm z}^L\sigma_{\rm z}^L$ uses $2n$ {\sc cphase} gates and a measurement of $\sigma_{\rm z}^L\sigma_{\rm z}^L\sigma_{\rm z}^L$ uses $3n$ {\sc cphase} gates. Therefore, the {\sc cnot} gadget contains $(2r_1+3r_2)n$ {\sc cphase} gates. We pessimistically assume that any non-dephasing fault in a {\sc cphase} gate that is either contained within the {\sc cnot} gadget or that propagates into the {\sc cnot} gadget causes the gadget to fail. We denote by $\varepsilon^{(1)}_{\neg d}$ the probability of failure due to a non-dephasing fault in a {\sc cphase} gate and conclude that
\begin{equation}
\varepsilon^{(1)}_{\neg d} \leq (2r_1 + 3r_2 + 2r)\, n \, \varepsilon' \; ,
\end{equation}
where $r\leq\max(r_1,r_2)$. Here, for each input block, $rn$ is an upper bound on the number of {\sc cphase} gates in the preceding gadget where errors can propagate into the current gadget.

Now, suppose that there are no {\sc cphase} gates with non-dephasing faults, and consider the probability of failure of the {\sc cnot} gadget due to dephasing faults at {\sc cphase} gates, and due to faults in preparations and measurements. We may assume without loss of generality that the faults at the operations $\mathcal{P}_{|+\rangle}$ and $\mathcal{M}_{\sigma_{\rm x}}$ have diagonal Kraus operators, as $\sigma_{\rm x}$ errors have no effect at these locations.

A dephasing fault can alter the outcome of a $\sigma_{\rm z}^L\sigma_{\rm z}^L$ or $\sigma_{\rm z}^L\sigma_{\rm z}^L\sigma_{\rm z}^L$ measurement only if it acts on the ancilla qubit used during the measurement. For each of these logical measurements there is one preparation and one measurement operation; furthermore, there are $2n$ {\sc cphase} gates for the $\sigma_{\rm z}^L\sigma_{\rm z}^L$ measurement and $3n$ {\sc cphase} gates for the $\sigma_{\rm z}^L\sigma_{\rm z}^L\sigma_{\rm z}^L$ measurement. We therefore obtain upper bounds on the probability of failure for the majority vote of the repeated measurements:
\begin{eqnarray}
\label{failure-of-measurement}
\varepsilon(\mathcal{M}_{\sigma_{\rm z}^L \sigma_{\rm z}^L}) \leq {r_1 \choose  {r_1+1\over 2} } \left(( 2n+2) \, \varepsilon \right)^{ {r_1+1 \over 2} } \; , \\ 
\varepsilon(\mathcal{M}_{\sigma_{\rm z}^L \sigma_{\rm z}^L \sigma_{\rm z}^L}) \leq {r_2 \choose  {r_2+1\over 2} } \left(( 3n+2) \, \varepsilon \right)^{ {r_2+1 \over 2} } \; .
\end{eqnarray}

A measurement of $\sigma_{\rm x}^L$ can fail if the majority of the qubits in the block have errors. For each qubit, the error can arise during the preparation of the qubit, the measurement of the qubit, or a {\sc cphase} gate that acts on the qubit. Therefore, an upper bound on the probability of failure is 
\begin{equation}
\varepsilon(\mathcal{M}^{[1]}_{\sigma_{\rm x}^L}) \leq {n \choose {n+1\over 2} } \left( (r + r_1 + r_2 +2) \, \varepsilon\right)^{ {n+1 \over 2} } \; 
\end{equation}
for the measurement of the control block, and 
\begin{equation}
\varepsilon(\mathcal{M}^{[2]}_{\sigma_{\rm x}^L}) \leq {n \choose  {n+1\over 2} } \left( ( r + r_2+2 ) \, \varepsilon \right)^{ {n+1 \over 2} } \; 
\end{equation}
\noindent for the measurement of the target block. Denoting by $\varepsilon^{(1)}_{d}$ the probability of failure due to faults other than non-dephasing faults in {\sc cphase} gates, we obtain
\begin{eqnarray}
\varepsilon^{(1)}_{d} &\leq& \varepsilon(\mathcal{M}_{\sigma_{\rm z}^L \sigma_{\rm z}^L}) + \varepsilon(\mathcal{M}_{\sigma_{\rm z}^L \sigma_{\rm z}^L \sigma_{\rm z}^L}) + \varepsilon(\mathcal{M}^{[1]}_{\sigma_{\rm x}^L}) + \varepsilon(\mathcal{M}^{[2]}_{\sigma_{\rm x}^L}) \nonumber\\
& \leq & 2{n \choose  {n+1\over 2} } \left[( 2n+2)^{ {n+1 \over 2} } + ( 3n+2)^{ {n+1 \over 2}}\right] \varepsilon ^{ {n+1 \over 2} } \; ,
\end{eqnarray}
\noindent where to obtain the second inequality we have substituted $r_1=r_2=r=n$. Our upper bound on the total probability of failure for the {\sc cnot} gadget is
\begin{equation}
\varepsilon^{(1)}\le \varepsilon^{(1)}_{\neg d}+\varepsilon^{(1)}_{d} \le 7n^2\varepsilon' + \varepsilon^{(1)}_{d}~.
\end{equation}

The quantity $\varepsilon^{(1)}$ is the effective noise strength for our $\mathcal{C}_1$-protected $\mathcal{G}_{\rm CSS}$ gadgets. In particular, if we set $r=n=11$, $\varepsilon = 2.50\times 10^{-3}$, and $\varepsilon/\varepsilon' = 10^{4}$, we find $\varepsilon^{(1)} < .67\times 10^{-3}$ so that the effective noise strength is below the threshold $\varepsilon^{\rm CSS}_{\rm th} \geq .67\times 10^{-3}$ for $\mathcal{G}_{\rm CSS}$ gadgets protected by $\mathcal{C}_2$ \cite{fibonacci}. Thus $2.50\times 10^{-3}$ is a lower bound on the accuracy threshold for $\mathcal{G}_{\rm CSS}$ operations assuming a local stochastic biased noise model with bias $10^4$. 



\section{Accuracy threshold for universal quantum computation}

\begin{figure}[t]
\begin{center}
\begin{tabular}{c}
\put(-2.1,0){\includegraphics[width=15.5cm,keepaspectratio]{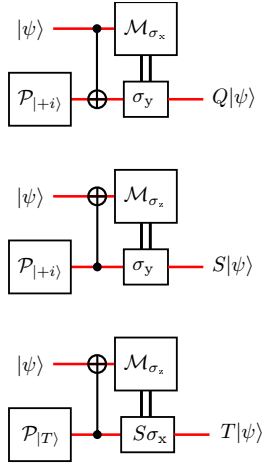}}
\vspace{-5.7cm}
\end{tabular}
\end{center}\caption{Teleportation circuits for the Clifford group gates $Q=e^{i(\pi/4)\sigma_{\rm x}}$ and $S=e^{-i(\pi/4)\sigma_{\rm z}}$, and for the non-Clifford gate $T=e^{-i(\pi/8)\sigma_{\rm z}}$. For the Clifford group gates the measurement determines a Pauli operator that updates the Pauli frame, and for the non-Clifford gate the measurement determines a non-Pauli correction that must be applied in the next step.}
\label{fig:gate-teleportation}
\end{figure}

If $\varepsilon$ is below the $\mathcal{G}_{\rm CSS}$ threshold, $\mathcal{G}_{\rm CSS}$ gadgets protected by $\mathcal{C}_1 \triangleright \mathcal{C}_2$ are highly reliable. To extend our gadgets to a universal set protected by $\mathcal{C}_1 \triangleright \mathcal{C}_2$, we use gate teleportation as shown in Fig.~\ref{fig:gate-teleportation}. Provided we can prepare the state $|{+}i\rangle$, we can use the $\mathcal{G}_{\rm CSS}$ operations {\sc cnot}, $\mathcal{P}_{|+\rangle}$, $\mathcal{M}_{\sigma_{\rm x}}$, and $\mathcal{M}_{\sigma_{\rm z}}$ to teleport $Q=\exp(i(\pi/4)\sigma_{\rm x})$ and $S=\exp(-i(\pi/4)\sigma_{\rm z})$. Together with the {\sc cnot} gate, $Q$ and $S$ suffice to generate the Clifford group. Provided we can prepare the state $|T\rangle$, we can go beyond the Clifford group and achieve universality by using $\mathcal{G}_{\rm CSS}$ operations and $S$ to teleport the gate $T=\exp(-i(\pi/8)\sigma_{\rm z})$. Thus, we can do reliable universal quantum computation if we can perform CSS operations reliably and we can also prepare high-fidelity copies of the state $|{+}i\rangle$ (the eigenstate of $\sigma_{\rm y}$ with eigenvalue +1) and the state $|T\rangle$ (the eigenstate of $S{\sigma_{\rm x}}$ with eigenvalue +1).

Furthermore, if we are able to prepare noisy copies of $|{+}i\rangle$ and $|T\rangle$ that are not {\em too} noisy, then high fidelity copies can be generated via state distillation protocols \cite{bravyi}. These protocols are based on CSS stabilizer codes, for which $\mathcal{G}_{\rm CSS}$ operations suffice to measure the error syndrome and to decode. The distillation protocol for $|{+}i\rangle$ uses Steane's [[7,1,3]] CSS code. In each round of the protocol, the code's check operators are measured for seven noisy copies of the input state; the encoded qubit is accepted and decoded if the error syndrome is trivial. The state $|{+}i\rangle$ is prepared successfully unless at least three of the input states have errors. Similarly, the distillation protocol for $|T\rangle$ uses a [[15,1,3]] CSS code. In each round of the protocol, the code's check operators are measured for fifteen noisy copies of the input state; the encoded qubit is accepted and decoded if the error syndrome is trivial. Here, too, the state $|T\rangle$ is prepared successfully unless at least three of the input states have errors.

The error threshold for the $|T\rangle$ distillation protocol was estimated in \cite{bravyi}, where it was shown that an input error probability as high as $14.1\%$ can be tolerated if each input state is ``twirled'' by applying $S{\sigma_{\rm x}}$ with probability $1/2$. The error threshold for $|{+}i\rangle$ distillation is even higher. Therefore, if $\mathcal{G}_{\rm CSS}$ gadgets protected by $\mathcal{C}_1 \triangleright \mathcal{C}_2$ are reliable, and we can also inject input states into the $\mathcal{C}_1 \triangleright \mathcal{C}_2$ block with probability of error below $14.1\%$, then we can do reliable universal quantum computation. (By distilling $|{+}i\rangle$, we can teleport the $S$ gate, enabling us to perform the twirling step in the $|T\rangle$ distillation protocol.)

\begin{figure}[t]
\begin{center}
\begin{tabular}{c}
\put(-3.1,0){\includegraphics[width=17cm,keepaspectratio]{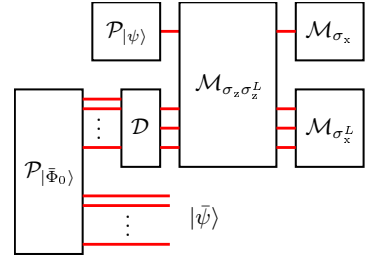}}
\vspace{-9.5cm}
\end{tabular}
\end{center}\caption{Injection of the state $|\psi\rangle$ into the $\mathcal{C}_1 \triangleright \mathcal{C}_2$ block by using teleportation. After the encoded Bell state $|\bar{\Phi}_0\rangle$ is prepared, one $\mathcal{C}_1 \triangleright \mathcal{C}_2$ block is decoded to $\mathcal{C}_1$. Then, a Bell measurement is performed on the $\mathcal{C}_1 $ block and the input state $|\psi\rangle$.}
\label{fig:inject-by-teleportation}
\end{figure}

The state injection is performed by teleportation as in Fig.~\ref{fig:inject-by-teleportation}. Let us use $|\bar \psi\rangle$ to denote a state encoded in $\mathcal{C}_1 \triangleright \mathcal{C}_2$, to distinguish it from $|\psi\rangle_L$, the state encoded in $\mathcal{C}_1$. To inject the single-qubit state $|\psi\rangle$ into the $\mathcal{C}_1 \triangleright  \mathcal{C}_2$ block, first the encoded Bell state $|\bar{\Phi}_0\rangle = {1\over \sqrt{2}}(|\bar{0}\rangle |\bar{0}\rangle + |\bar{1}\rangle |\bar{1}\rangle)$ is prepared, and then one of the code blocks is decoded to $\mathcal{C}_1$. To complete the teleportation, a joint Bell measurement is performed on the $\mathcal{C}_1$ block and the unprotected state $|\psi\rangle$. This procedure prepares the state $|\bar\psi\rangle$, up to a logical Pauli operator that is known from the outcome of the Bell measurement. 

Because $|\bar\Phi_0\rangle$ can be prepared by using $\mathcal{G}_{\rm CSS}$ gadgets which are well protected by $\mathcal{C}_1 \triangleright \mathcal{C}_2$ (we start with the encoded state $|\bar{+}\rangle$ and the encoded state $|\bar{0}\rangle$, and we apply an encoded {\sc cnot} gate), we may assume that the preparation of $|\bar\Phi_0\rangle$ is flawless. Thus the state injection might fail because of a decoding error, because of an error in the Bell measurement, or because of a fault during the single-qubit preparation of $|\psi\rangle$. 

Now let us suppose that $\mathcal{C}_2=\mathcal{C}^{\triangleright k}$ is obtained by concatenating the CSS code $\mathcal{C}$ all together $k$ times. The decoding of $\mathcal{C}_1\triangleright \mathcal{C}_2$ is performed recursively: In the first step, $\mathcal{C}_1\triangleright\mathcal{C}^{\triangleright k}$ is decoded to $\mathcal{C}_1\triangleright\mathcal{C}^{\triangleright (k-1)}$ using gadgets protected by $\mathcal{C}_1\triangleright\mathcal{C}^{\triangleright (k-1)}$; then,  $\mathcal{C}_1\triangleright\mathcal{C}^{\triangleright (k-1)}$ is decoded to $\mathcal{C}_1\triangleright\mathcal{C}^{\triangleright (k-2)}$ using gadgets protected by $\mathcal{C}_1\triangleright\mathcal{C}^{\triangleright (k-2)}$, and so on. In the last step, $\mathcal{C}_1\triangleright\mathcal{C}$ is decoded to $\mathcal{C}_1$ using gadgets protected by $\mathcal{C}_1$. 
%
Let us denote by $\varepsilon(\mathcal{D})$ the probability that a logical error occurs at any level during this recursive decoding. If decoding is staggered so that no qubits are idle during the Bell measurement, and if the measurement of $\sigma_{\rm z}\sigma_{\rm z}^L$ is repeated $r$ times inside $\mathcal{M}_{\sigma_{\rm z}\sigma_{\rm z}^L}$, the probability of a state injection error is 
\begin{equation}
\label{eq:C1}
\varepsilon(\mathcal{P}_{|\bar \psi\rangle}) \leq \varepsilon(\mathcal{D}) + \varepsilon_{\rm BM} + \varepsilon ~,
\end{equation}
\noindent where $\varepsilon$ accounts for the probability of a fault in the single-qubit preparation of $|\psi\rangle$ and
\begin{equation}
\begin{array}{rcl}
\varepsilon_{\rm BM} & \le & (2rn+r)\varepsilon' + (1+r)\varepsilon \\
                     &     & + {r \choose {r+1\over 2} } ((n+3) \varepsilon)^{r+1\over 2} + {n \choose {n+1\over 2} } ((2r+2)\varepsilon)^{n+1\over 2} \; . 
\end{array}
\label{ancilla-error}
\end{equation}
\noindent 
Here, $(2rn{+}r)\varepsilon'$ bounds the probability of error in the Bell measurement due to a non-dephasing fault in a {\sc cphase} gate; the fault could occur in one of the $r(n{+}1)$ gates contained in $\mathcal{M}_{\sigma_{\rm z}\sigma_{\rm z}^L}$, or in one of the $rn$ gates contained in a measurement in the immediately preceding $\mathcal{C}_1$-protected {\sc cnot} gadget (which is part of the recursive decoding circuit). Furthermore, $(1{+}r)\varepsilon$ bounds the probability of error in  $\mathcal{M}_{\sigma_{\rm x}}$ due to a dephasing fault; the measured qubit participates in $r$ {\sc cphase} gates contained in $\mathcal{M}_{\sigma_{\rm z}\sigma_{\rm z}^L}$ and also in the measurement itself (the probability of a fault in the preparation of this qubit in the state $|\psi\rangle$ has already been included in Eq.~(\ref{eq:C1})). The next to last term bounds the probability of error in $\mathcal{M}_{\sigma_{\rm z}\sigma_{\rm z}^L}$ due to a dephasing fault; the ancilla qubit in each of the $r$ measurements inside $\mathcal{M}_{\sigma_{\rm z}\sigma_{\rm z}^L}$ participates in one $\mathcal{P}_{|+\rangle}$, $n{+}1$ {\sc cphase} gates, and one $\mathcal{M}_{\sigma_{\rm x}}$. Finally, the last term bounds the probability of an error in $\mathcal{M}_{\sigma_{\rm x}^L}$; each qubit in the measured block participates in one $\mathcal{P}_{|+\rangle}$, $r$ {\sc cphase} gates contained in the preceding $\mathcal{C}_1$-protected {\sc cnot} gadget, $r$ {\sc cphase} gates contained in $\mathcal{M}_{\sigma_{\rm z}\sigma_{\rm z}^L}$, and one $\mathcal{M}_{\sigma_{\rm x}}$.

If we set $r=n= 11$, $\varepsilon \leq 2.50\times 10^{-3}$, and $\varepsilon' \leq 10^{-4}\varepsilon$, we find $\varepsilon_{\rm BM} \leq 3.01 \%$. Since $\varepsilon(\mathcal{D})\leq 8.24\%$ for $\varepsilon^{(1)}\leq .67\times 10^{-3}$ \cite{fibonacci}, we conclude that $\varepsilon(\mathcal{P}_{|\bar \psi\rangle}) \leq 11.5\%$ which is below the $14.1\%$ distillation threshold. Thus $2.50\times 10^{-3}$ is a lower bound on the accuracy threshold for universal quantum computation under local stochastic biased noise model with bias $10^4$. 


\section{Improved threshold via flagging and message passing}

We can improve our lower bound on the accuracy threshold by using a more sophisticated decoding procedure for $\mathcal{C}_1 \triangleright \mathcal{C}_2$. We note that the syndrome information for $\mathcal{C}_1$ is helpful for optimizing the decoding of $\mathcal{C}_2$ in the concatenated block; yet the decoding procedure that we have described so far makes no use of this information---after $\mathcal{C}_1$ is decoded, the $\mathcal{C}_1$-syndrome is discarded. Now we consider a new decoding procedure where some information about the $\mathcal{C}_1$-syndrome is retained and used in the decoding of $\mathcal{C}_2$. 

For the sake of clarity, we continue to make a distinction between $r_1$, $r_2$ and $n$, even though we will set them equal later on. We say that a vote is ``close'' if the winners have one more vote than the losers. Thus $\mathcal{M}_{\sigma_{x}^L}$ has a close vote if the error syndrome indicates $(n{-}1)/2$ errors in the block, $\mathcal{M}_{\sigma_{z}^L\sigma_{\rm z}^L}$ has a close vote if $(r_1{-}1)/2$ of the measurements disagree with the majority, and  $\mathcal{M}_{\sigma_{z}^L\sigma_{\rm z}^L\sigma_{\rm z}^L}$ has a close vote if $(r_2{-}1)/2$ measurements disagree with the majority. If a gadget contains no close votes, then we decode $\mathcal{C}_1$ as usual. But if the gadget contains a close vote, then a {\em flag} is raised after decoding. The flag signifies that the gadget has a higher than usual probability of failure, information that will be exploited during decoding at the next level up in the concatenated block, using a scheme described in \cite{knill}.

\begin{figure}[t]
\begin{center}
\begin{tabular}{c}
\put(-3.6,0){\includegraphics[width=17cm,keepaspectratio]{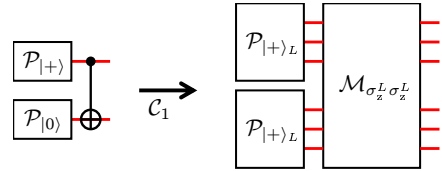}}
\vspace{-10.7cm}
\end{tabular}
\end{center}\caption{A Bell state $|\Phi_0\rangle_L$ is prepared by starting with two blocks in the state $|+\rangle_L$ and then measuring $\sigma_{\rm z}^L \sigma_{\rm z}^L$.}
\label{fig:bell-pair}
\end{figure}

\begin{figure*}[t]
\begin{center}
\begin{tabular}{c}
\put(-5,-1){\includegraphics[width=14cm,keepaspectratio]{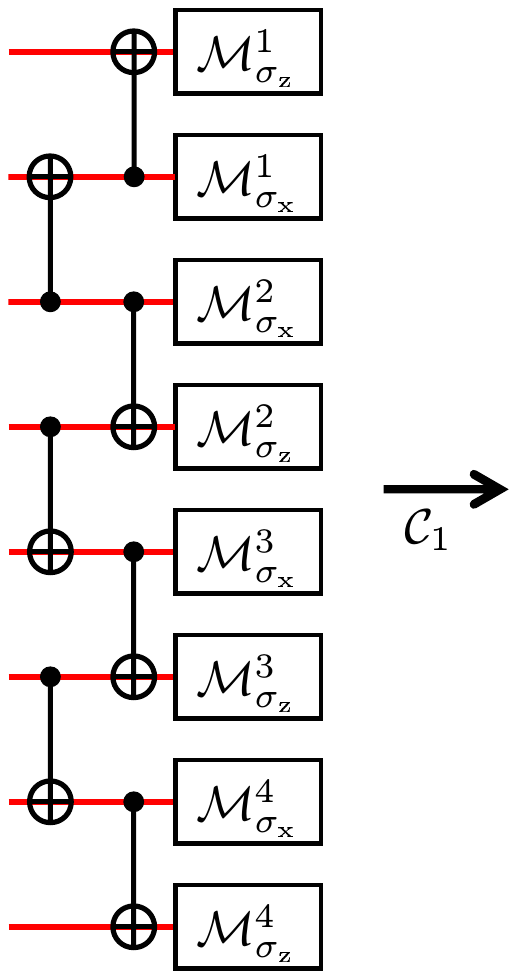}}
\put(-1.8,0){\includegraphics[width=14.3cm,keepaspectratio]{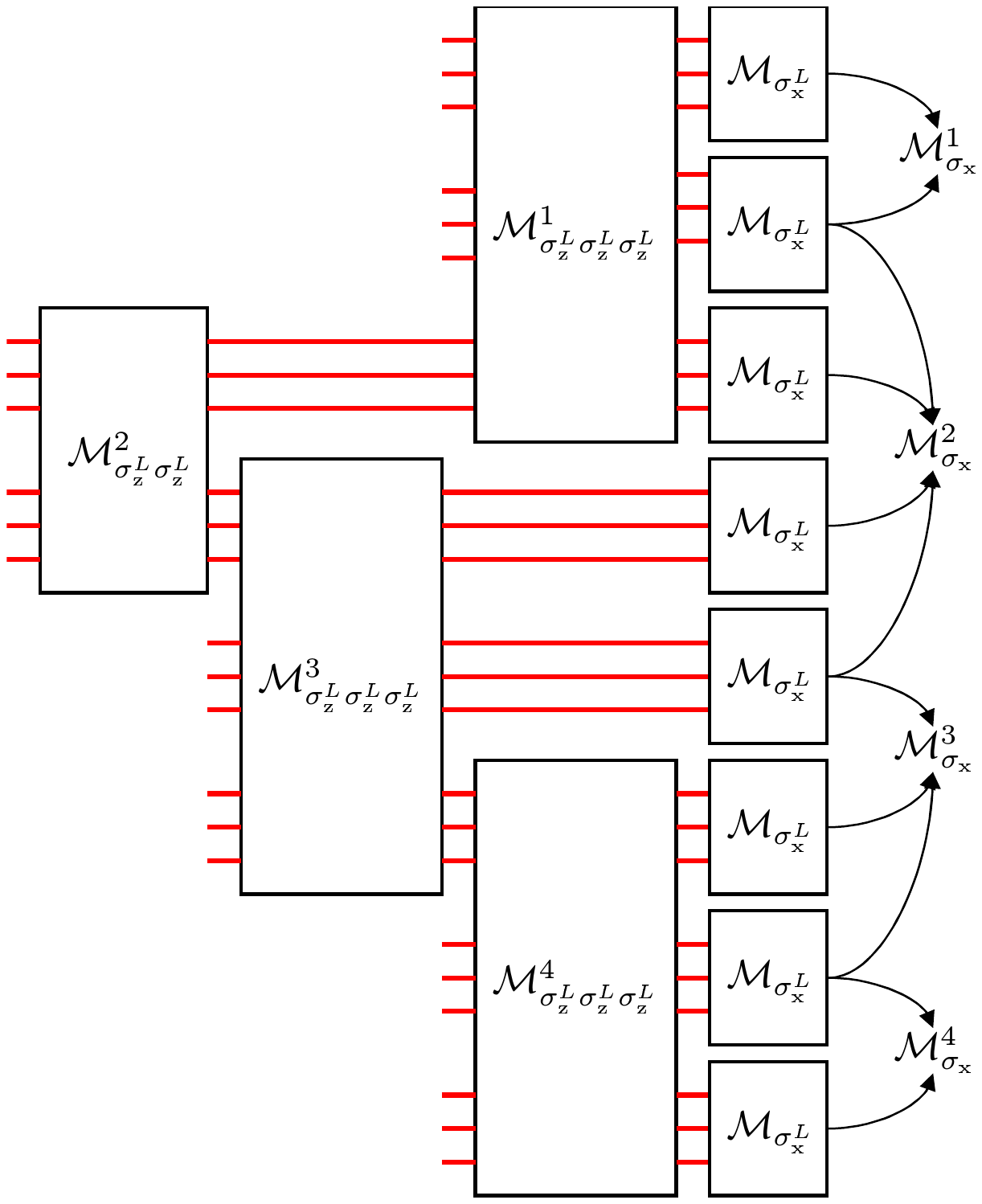}}
\vspace{-4.1cm}
\end{tabular}
\end{center}\caption{On the left, the {\sc cnot} gates and measurements used to implement the teleportations of $\mathcal{C}_2$ blocks and subblocks in Fig.~3 in \cite{fibonacci}. On the right, a $\mathcal{C}_1$-protected implementation of the operations on the left. The outcome of a measurement of $\sigma_{\rm z}^L\sigma_{\rm z}^L$ or $\sigma_{\rm z}^L\sigma_{\rm z}^L\sigma_{\rm z}^L$ on the right corresponds to the outcome of the measurement of $\sigma_{\rm z}$ on the left with the same superscript, while the outcomes of the measurements of $\sigma_{\rm x}^L$ on the right are combined as shown to give the outcomes of the measurements of $\sigma_{\rm x}$ on the left.}
\label{fig:bell-measurements}
\end{figure*}

For simplicity, we consider a version of the scheme in \cite{knill} where $\mathcal{C}_2$ is the concatenated 4-qubit code ($C_4$) with check operators $\sigma_{\rm x}^{\otimes 4}$ and $\sigma_{\rm z}^{\otimes 4}$; this is the case analyzed in \cite{fibonacci}. The basic building blocks for the construction of $\mathcal{C}_2$-protected $\mathcal{G}_{\rm CSS}$ operations in this scheme are Bell states $|\Phi_0\rangle = {1\over \sqrt{2}}(|0\rangle |0\rangle + |1\rangle |1\rangle)$ as on the left of Fig.~\ref{fig:bell-pair},  and {\sc cnot} gates followed by single-qubit measurements as on the left of Fig.~\ref{fig:bell-measurements}. We can then construct encoded versions of these two basic building blocks by using $\mathcal{C}_1$-protected gadgets as on the right of Figs.~\ref{fig:bell-pair} and \ref{fig:bell-measurements}.

Consider how to refine our estimate of the failure probability for $\mathcal{M}_{\sigma_{z}^L\sigma_{\rm z}^L}$ in Fig.~\ref{fig:bell-pair} taking flagging into account. If there is a flag, then $(r_1{+}1)/2$ faulty measurements might cause $\mathcal{M}_{\sigma_{z}^L\sigma_{\rm z}^L}$ to fail, but without a flag at least $(r_1{+}3)/2$ faulty measurements are required. Therefore, instead of Eq.~(\ref{failure-of-measurement}) we have
\begin{equation}
\varepsilon_{f}(\mathcal{M}_{\sigma_{\rm z}^L\sigma_{\rm z}^L}) \leq {r_1 \choose {r_1+1 \over 2}} \left((2n+2)\varepsilon \right)^{{r_1+1 \over 2}}  
\end{equation}
\noindent for the probability of failure with a flag, and
\begin{equation}
\varepsilon_{\neg f}(\mathcal{M}_{\sigma_{\rm z}^L\sigma_{\rm z}^L}) \leq {r_1 \choose  {r_1+3 \over 2} } \left((2n+2)\varepsilon \right)^{ {r_1+3 \over 2}} 
\end{equation}
\noindent for the probability of failure without a flag. In fact, because the circuit in Fig.~\ref{fig:bell-pair} prepares an ancilla state, we find it advantageous to repeat this measurement of $\sigma_{z}^L\sigma_{\rm z}^L$ a smaller number of times $t< n$ and to postselect on the cases without a flag. In this case, the conditional probability that $\mathcal{M}_{\sigma_{z}^L\sigma_{\rm z}^L}$ fails when it is accepted is 
\begin{equation}
\varepsilon ( \mathcal{M}_{\sigma_{\rm z}^L\sigma_{\rm z}^L} | {\rm acc} ) \leq { {t \choose  {t+3 \over 2} } \left((2n+2)\varepsilon \right)^{ {t+3 \over 2}} + 2nt\varepsilon' \over 1 - {t \choose {t-1 \over 2}} \left((2n+2)\varepsilon \right)^{{t-1 \over 2}} - 2nt\varepsilon' } \; , 
\end{equation}
\noindent where ${t \choose {t-1 \over 2}} \left((2n+2)\varepsilon \right)^{{t-1 \over 2}}$ bounds the probability that a flag is raised, and $2nt\varepsilon'$ is the probability of a non-dephasing error in one of the {\sc cphase} gates. 

We can perform a similar analysis to bound the probability of failure, with and without a flag, for the measurements $\mathcal{M}_{\sigma_{z}^L\sigma_{\rm z}^L}$, $\mathcal{M}_{\sigma_{z}^L\sigma_{\rm z}^L\sigma_{\rm z}^L}$, and $\mathcal{M}_{\sigma_{x}^L}$ in Fig.~\ref{fig:bell-measurements}. These bounds can now be plugged into the analysis of the decoding of $\mathcal{C}_2$ in \cite{fibonacci}. For bias $\varepsilon/\varepsilon' = 10^{4}$, we find that reliable $(\mathcal{C}_1 \triangleright \mathcal{C}_2)$-protected $\mathcal{G}_{\rm CSS}$ operations can be implemented if $\varepsilon$ is below $3.51\times10^{-3}$ (where this optimal value is achieved by choosing $r_1=r_2=n=7$ and $t=5$). In addition, by an analysis similar to the discussion in Appendix C, we find that for $\varepsilon=3.51 \times 10^{-3}$ the probability of an error in state injection is $\varepsilon(\mathcal{P}_{|\bar{\psi}\rangle}) \leq 10.4\%$, which is below the $14.1\%$ distillation threshold. Thus $3.51\times10^{-3}$ is our improved lower bound on the accuracy threshold for universal quantum computation under local stochastic biased noise with bias $10^4$. 

\end{document}